\documentstyle[11pt,epsfig,amssymb]{article}
\newlength{\dinwidth}
\newlength{\dinmargin}
\newcommand {\nn} {\nonumber}
\newcommand {\half} {\frac{1}{2}}
\newcommand {\p} {\prime}

\renewcommand{\theequation}{\arabic{equation}}

\begin{document}
\topmargin -0.8cm
\oddsidemargin -0.8cm
\evensidemargin -0.8cm
\pagestyle{empty}
\begin{flushright}
SNUTP-01-011\\
{\tt hep-th/0104226}
\end{flushright}
\centerline{\Large \bf Gravitational potential correction with} 
\vskip0.4cm
\centerline{\Large \bf Gauss-Bonnet interaction}
\vspace*{0.8cm}
\centerline{\bf Ishwaree P. Neupane\,
\footnote{Email: ishwaree@phya.snu.ac.kr}}
\vskip.1in
\centerline{\it School of Physics \& Center for Theoretical Physics}
\vspace{1mm}
\centerline{\it Seoul National University, Seoul, 151-742, Korea}
\vspace*{1.0cm}
\centerline{\bf Abstract}
\vspace*{0.4cm}
A treatment of linearized gravity with Einstein-Gauss-Bonnet interaction 
terms in $D\geq5$ is given in the Randall-Sundrum brane background. This 
Letter has outlined some interesting features of the brane world gravity and 
Newtonian potential correction with GB interaction term. We find that the 
GB coupling $\alpha$ renormalizes the effective four-dimensional Newton 
constant on the brane, and also additionally contributes to the correction 
term of the Newtonian potential. Indeed, the GB term does not affect 
the massless graviton mode and the Einstein gravity on the brane , and quite 
interestingly, such term in $D\geq 5$ appears to give more information about 
the necessary boundary condition(s) to be satisfied by the zero mode 
wavefunction on the brane(s).   
\begin{flushleft}
\end{flushleft}
\pagestyle{plain}
\section{Introduction and overview}
Brane world scenarios are presently the subject of entensive 
investigation. This is partly due to some success of 
ArkaniHamed-Dimopoulos-Dvali (ADD) approach~\cite{ADD} to tackle the mass 
hiearchy and small four-dimensional cosmological constant problems by 
reducing the mass scale from the Planck scale $\sim 10^{19} GeV$ to the 
fundamental $TeV$ scale. This is done by using a large but compact extra 
dimension, which leads to possibilities for the size of extra space 
$\sim 1 mm$. There is also intense interest in the 
Randall-Sundrum brane world proposal~\cite{RS}, wherein a non-compact extra 
dimension is used to localize gravity on a $3$-brane, which is embedded in 
an $AdS_5$ bulk. The ideas 
behind a non-factorizable geometry of the space-time induced by a 
strongly gravitating $3$-brane lead to interesting consequences for black 
hole~\cite{CHR}, global $p$-brane~\cite{MRK} and $D3$ brane~\cite{BCM} 
solutions in some extension of the earlier proposals. It has been argued 
in~\cite{BCM} that the apparent infinite extension of the interior 
$AdS$ region could be due to the warping of the extra dimensions by 
gravity of the $D3$ brane. The RS proposals have also resulted in new work 
on brane world Kaluza-Klein reductions~\cite{MCV} and various others in 
different directions. 

At the linearized level [2, 7-12], Einstein gravity can be combined with 
a warped geometry in $5d$, a positive tension (vacuum energy) $3$-brane 
and a negative cosmological constant in bulk $AdS_5$, and the 
fluctuations of the $4d$ metric are described by an analog quantum 
mechanical Schr\"odinger equation with an attractive delta function 
potential generated by the $AdS$ curvature scale. The eigen 
spectrum supports precisely one bound state with zero energy, which is 
identifed as the $4d$ massless graviton, along with suppressed 
wave functions of the massive gapless KK states near the brane. Indeed, the 
negligible KK modes for small energy implies the existence of 
ordinary $4d$ gravity at large distances on the brane. A common 
assumption in compactified KK theories \cite{ACF} is that with a small 
periodic fifth dimension, the low-energy degrees of freedom are to be 
restricted to zero modes, well separated by a large mass gap from the 
massive modes, and that these modes be independent of the internal 
coordinate(s). By contrast, the RS type compactfication scheme may look 
unconventional and non-standard~\cite{BKS}, not only from the above 
arguments, but also because the coupling of matter to 
gravity appears through Neumann-type boundary condition instead of 
Einstein equations on the boundary \cite{SMS}. In particular, a 
four-dimensional low-energy effective field theory does not follow from the 
usual KK reasoning. Nonetheless, RS gravity on the boundary ($3$-brane) is 
appealing, as it reproduces the correct Newtonian limit and the zero-mode 
truncation of the $5d$ theory coincides with the usual linearized 
$(3+1)$-dimensional gravity.

In this regard, a natural option on the route to a consistent brane world 
gravity in $D\geq 5$ is to introduce higher-curvature terms to the 
Einstein-Hilbert action, because the effective $4d$ gravity on $3$-brane 
should admit such terms. These should be derived from the low energy 
effective action of string theory, and introduced as a ghost-free 
Gauss-Bonnet (GB) combination [16-20]. The generic form of higher-curvature 
terms in the brane background tend to delocalize gravity~\cite{CZK} on the 
brane. However, such terms in the GB combination can result in a localized 
gravity at long distances along the brane. 

Interestingly, the massless modes $(m^2=0)$ of the graviton 
wavefunctions remain unaffected in the presence of GB term. In particular, 
with the GB term, the effective volcano-type potential $(V_{eff.})$ in 
$D\geq 5$ involve extra terms (or in combination) which are trivial for a 
normalizable zero mode wavefunction due to the necessary boundary 
condition(s) to be satisfied on the brane(s). As expected, the GB 
coupling redefines the effective $4d$ Newton constant on the brane, and 
also additionally contributes to the leading order correction to the 
Newtonian potential. The former property entails that the GB term does not 
change the Einstein gravity on the $3$-brane, but, meanwhile, can give 
further information about the boundary condition(s) on the brane(s). Further, 
a brane world model in $D\geq 5$ appears to shed more light 
toward the embedding of the RS scenario into a general $AdS_D$ 
background, and this may also be realized in some setting of supergravity 
theories. Indeed, to 
regard the brane tension of intersecting $n+2$ branes as the true vacuum 
energy of the visible world, one may have to introduce the higher-curvature 
terms, because such terms even in pure gravitational $AdS$ background, 
possibly in GB combination, can support feasible singularities produced at 
the brane and hence allow a non-trivial brane tension or four-dimensional 
cosmological constant. This is indeed required for a meaningful realization 
of the RS model in $D>5$. In this Letter we will be more confined to $D=5$ 
case, and point out briefly some important features in $D\geq 6$ at the end.   
\section{Background with Poincar\'e symmetry}
A general $D (=d+1)$-dimensional metric background possessing $4d$ 
Lorentzian symmetry is given by
\begin{equation}
ds^2= e^{-2A(z)} g_{\mu\nu}(x^\lambda)dx^{\mu}dx^{\nu}
+g_{ij}(z)dz^i dz^j= g_{ab}dx^a dx^b
\label{genmetric}\,.
\end{equation}
Here $x^a=(x^{\mu},z^i)$, where $x^{\mu}(\mu = 0,...,3)$ are the usual 
coordinates of the Lorenzian space, and $z^i (i= 5,...,d+1)$ are the 
coordinates of the $(D-4)$-dimensional transverse space. 
With $g_{\mu\nu}(x^\lambda)=\eta_{\mu\nu}$, 
the metric ~(\ref{genmetric}) in $D=5$ translates to a more relevance form   
\begin{equation}
ds^2= e^{-2A(y)}\eta_{\mu\nu}dx^{\mu}dx^{\nu}+dy^2={\hat g}_{ab}dx^a dx^b\,.
\label{rsmetric}
\end{equation}
The transverse coordinate $y$ here defines a preferred family of 
$y=const$ hypersurfaces. An arbitrary metric of the 
form~(\ref{genmetric}) can be 
brought to the following conformally flat metric, using 
$z=\partial_y|y|\,\ell\,(e^{|y|/\ell}-1)$,    
\begin{equation}
ds^2= e^{-2A(z_i)}(\eta_{\mu\nu}dx^{\mu}dx^{\nu} + dz_i^2)
={\bar g}_{ab}dx^a dx^b\,,
\label{conmetric}
\end{equation}
where $A(y)=|y|/\ell$ is mapped to $A(z)= \log (|z|/\ell+1)$, and $\ell$ is 
the $AdS$ curvature scale\footnote{In $D(=4+n)$-dimensions with intersecting 
$n+2$-branes $A(z_i)$ can have a solution 
$A(z_i)=\log\big(|z_i|/\ell_i+1\big)$, where $z_i=z_1,\,z_2,\cdots$ count 
extra spaces and $\ell_i=\ell_1,\,\ell_2,\cdots$ are the $AdS$ curvature 
radii.}. In the conformal frame, the perturbed metric in $D=5$ is 
\begin{equation}
ds^2= e^{-2A(z)}\big\{(\eta_{\mu\nu}+h_{\mu\nu})dx^{\mu}dx^{\nu} + 
2h_{\mu 5} dx^\mu dz+(1+h_{55})dz^2\big\}
\label{perturbmet}\,.
\end{equation} 
The $Z_2$ symmetry $z\to -z$ implies that 
$h_{\mu 5}(x,z)=-h_{\mu 5}(x,-z)=0$ on the brane. It is also acceptable to 
argue from $4d$ point of view that the two physical 
degrees of freedom coming from the gravi-photon are not relevant for a 
matter-localized brane, so that one can gauge away the vector 
$h_{\mu 5}$ component everywhere, but the (gravi)scalar 
$h_{55}$ component, in general, on the basis of standard 
compacified KK theories, contributes to the physical processes. 
However, in the RS brane set up these fields do not correspond to the 
propagating degrees of freedom, and one can gauge away them, in 
particular, in the absence of matter source on the brane. 
\section{Effective action and linearized gravity}
We start with the following effective action in 
$5$-dimensional space-time $(M)$, where $\partial M$ represents the $4$-
dimensional boundary,   
\begin{eqnarray}
S&=&\int_{M} d^5x\,\sqrt{-g}\,\Big\{\kappa^{-1} R-2\Lambda
+\alpha \big( R^2- 4 R_{ab}R^{ab}
+ R_{abcd}R^{abcd}\big)\Big\}\nn\\
&&+\int_{\partial M}d^4\,x\sqrt{-\gamma}\,\big({\cal L}_m^{bdry}-V(z)\big)
 +\int_{M} d^5x\, \sqrt{-g} 
\,{\cal L}_{m}^{bulk}
\label{action}\,.
\end{eqnarray}
Here $\Lambda$ is the $5$-dimensional bulk cosmological constant, 
$V(z)$ is the vacuum energy (or tension) of the brane, and 
$\gamma_{\mu\nu}$ is the induced metric on the $y=const\, 4d$ hypersurface 
(or $3$-brane) and satisfies a completeness relation. 
The $D(=d+1)$ dimensional mass term is defined by 
$\kappa = 16 \pi G_{d+1} = M_*^{1-d}$, where $M_*$ is the 
$(d+1)$-dimensional fundamental mass scale. 
For a vacuum brane ${\cal L}_m^{bdry}=0$. The GB coupling $\alpha$ has the 
mass dimension of $M_*^{d-3}$, and in $D=5$ one defines 
$\alpha=M_* \alpha'$, where $\alpha'$ is the effective dimensionless 
coupling. In the RS scenario, $\ell$ takes an expected value of the order the 
fundamental (string) scale, 
$M_*\sim M_{pl}$ and $\ell M_*\sim 1$, however, in the ADD picture due to 
an expected large mass hierarchy, $\ell M_*>>1$ is implied. In both 
scenarios one basically works at energies below $M_{pl}$, so that the extra 
space can be effectively reduced to a one-dimensional space. 

First we set $\alpha=0$ in the equation~(\ref{action}). 
The $5d$ KK compactification tells us that the 
fluctuations $h_{55}$ and $h_{\mu 5}$ could represent, respectively, 
the real (gravi-)scalar and (gravi-)vector fields in the usual 
$(3+1)$-dimensions. 
Thus one may look at the possibility of a covariant prescription of the 
$5d$ harmonic (de-Donder) gauge, $\partial^a h_{ab} = 0 = h_a^a$. This 
implies that $h_\mu^\mu=-h_{55},\, \partial^\mu h_{\mu 5}=-\partial_5 
h_{55},\,\partial^\mu h_{\mu\nu}=-\partial_5 h_{5\nu}$. 
Consider the solution $A=\log (|z|/\ell+1)$ and the 
fine tuning conditions\footnote{This is just a reflection of the assumption 
that the $3$-brane world volume is Minkowskian. 
Under the RS choice $h_{ai}=0$ and $4d$ 
transverse-traceless (TT) gauge, one can define $\delta T^{(0)}_{\mu\nu}
=T_{\mu\lambda} h_\nu^\lambda$.} 
$\Lambda= - 12\kappa^{-1}/\ell^2$ and 
$ V = 12 \kappa^{-1}/\ell$. Then in the component form the linearlized 
version of the Einstein 
equations in the five dimensions take the form 
\begin{eqnarray}
&&\Big(\partial^2 + \frac{12}{(\ell+|z|)^2}+3f\partial_5\Big)h_{55}
=- \kappa T^{(m)}_{55}\label{h55compo}\\
&&\Big(\partial^2 + \frac{12}{\ell}\delta(z)\Big)h_{\mu 5}
+3f\partial_{\mu}h_{55}=-\kappa T^{(m)}_{\mu 5}\label{hmu5compo}\\
&&(\partial^2-3f\partial_5)h_{\mu\nu}+3f(\partial_{\mu}h_{\nu 5}
+\partial_{\nu} h_{\mu 5})+12\Big(\frac{1}{(\ell+|z|)^2}-\frac{\delta(z)}
{2\ell}
\Big)h_{55}\eta_{\mu\nu}=-\kappa T^{(m)}_{\mu\nu}\,,
\label{hmunucompo}
\end{eqnarray}
where $\partial^2=\eta^{ab}\partial_a\partial_b$ is the scalar $AdS$ 
Laplacian and $f(z)\equiv\partial_z A$. 
To study the vacuum propagation of metric fluctuations, one sets 
$T^{(m)}_{ab}=0$ and rescales $h_{55}$ in a canonically normalized form: 
$h_{55}(x,z)=e^{-3A(z)/2} {\tilde h}_{55}(x,z)$. Then one can look for 
solutions of the form ${\tilde h}_{55}(x,z)=\psi_5(z)\,{\tilde h}_{55}(x)$. 
Eq.~(\ref{h55compo}) with $\partial_\lambda^2{\tilde h}_{55}(x)
=m_5^2\,{\tilde h}_{55}$ then implies   
\begin{equation}
\Big[-\partial_z^2+V_5(z)\Big]\psi_5(z)=m_5^2\,\psi_5(z)\label{eqnv5}\,,
\end{equation}
where 
\begin{equation}
V_5(z)=\frac{3}{\ell} \delta(z) - \frac{45}{4(|z|+\ell)^2}\label{v5poten}\,.
\end{equation}
Because of the repulsive $\delta(z)$ term, it appears that the zero 
mode of the graviscalar cannot be localized on the brane. In fact, 
the consistency condition of eq.~(\ref{h55compo}) with 
eq.~(\ref{hmunucompo}) requires that $h_{55}$ (i.e., $\psi_5(z)$) stays 
zero at the brane. Further, consider the separation of variables 
$h_{\mu 5}(x,z)=\tilde{h}_{\mu 5}(x)\,\psi_v(z)$. 
Eq.~(\ref{hmu5compo}) with 
$\partial_\lambda^2{\tilde h}_{\mu5}
=m_v^2\,{\tilde h}_{\mu5}$ then leads to
\begin{equation}
\Big[-\partial_z^2+V_v(z)\Big]\psi_v(z)= m_v^2\,\psi_v(z)\label{vvpoten}\,,
\end{equation}
where 
\begin{equation}
V_v(z)=-\frac{12}{\ell} \delta(z)\,.
\end{equation}
This can admit a normalizable bound state solution. However, as it has no 
tail of the potential, $h_{\mu 5}$ cannot bear any physical 
(propagating) degrees of freedom. It appears that the 
normalizable bound-state solution 
$\psi_v(z)=~\sqrt{6/\ell}\, e^{-6|z|/\ell}$ admits a tachyonic mode. 
Notwithstanding this, a compatibility condition again with 
eq.~(\ref{hmunucompo}) implies that 
$h_{5\mu}=0\,\to \psi_v(0)=0$. This was adequately 
studied in~\cite{YSM}.
 
Finally, with the choice $h_{55}=0=h_{\mu 5}$ and 
defining $h_{\mu\nu}=e^{3A(z)/2}\epsilon_{\mu\nu} e^{ip.x}\psi(z)$, 
where $\epsilon_{\mu\nu}$ is the polarization tensor, one arrives at the 
following analog non-relativistic Schr\"odinger equation:     
\begin{equation}
\bigg[-\partial_z^2+ V_{eff}(z)\bigg]\psi(z)= m^2\psi(z)
\label{rsequation}\,.
\end{equation}
Here $V_{eff}$ is the RS potential, given by  
\begin{equation}
V_{eff}(z)=\frac{15}{4\big(|z|+\ell\big)^2}
-\frac{3}{\ell}\delta(z)
\label{rspotential}\,.
\end{equation}
The $m=0$ solution corresponds to a localized zero-mode graviton on the brane. 
In the RS scenario, the vector mode $h_{\mu 5}$ would appear with different 
behaviors than the scalar $(h_{55})$ and tensor $(h_{\mu\nu})$ modes. The 
Schr\"odinger equations for $h_{55}$ and $h_{\mu\nu}$ will have similar 
forms after the substitutions 
$\partial_z \to i\partial_z$ and $m \to i m_5$ in eq.~(\ref{rsequation}). 
This may hint that there do not exist any tachyonic bound state or 
(gravi)scalar modes in the vicinity of the RS brane. If any tachyonic scalar 
($M^2<0$) resides in the bulk, the RS background of a flat Minkowski 
$3$-brane would be unstable. Indeed, in order not to expect any tachyonic 
bound states of (gravi)scalar and vector modes , and to preserve the 
axial gauge $h_{a5}=0$ as well, one requires that the fluctuations $h_{55}$ 
and $h_{\mu 5}$ stay zero under some gauge transformations. 
\section{Gravitational potential with GB interaction}
We now study the tensor mode in $D=5$ by introducing higher-curvature terms 
in GB combination. The longitudinal components of 
the metric fluctuations are indeed the four-dimensional gravitons 
localized on the $3$-brane. With $4d$ TT condition and in axial gauge 
$h_{a 5}=0$, $h_{ab}$ only has non-zero components $h_{\mu\nu}$. Defining the 
symmetric stress tensor 
$\delta T^{(0)}_{\mu\nu}= {T^{(0)}}_{\mu\lambda}\, h_\nu^\lambda$, 
the linearized version of the field equations for $h_{\mu\nu}$ followed 
from~(\ref{action}) is given by 
\begin{equation}
(\partial^2-3f\partial_z)h_{\mu\nu}+4\alpha \kappa e^{2A(z)}
\Big\{\partial_zf \partial_\lambda^2-f^2\partial_z^2-f
\big(2\partial_zf-f^2\big)\partial_z\Big\}h_{\mu\nu}
= -\kappa T^{(m)}_{\mu\nu}
\label{linearized}\,,
\end{equation}
where $f(z)\equiv \partial_z A$. For $\alpha=0$, it reduces to 
the first bracket of eqn~(\ref{hmunucompo}), where all other terms in 
(~\ref{hmunucompo}) vanish due to the gauge choice. For a vacuum brane 
$T^{(m)}_{\mu\nu}=0$ and the RS background solution 
$A(z)=\log (|z|/\ell+1)$ in $D=5$, eq.~(\ref{linearized}) takes the form
\begin{eqnarray}
&\bigg[-\half\bigg(1-\frac{4\alpha\kappa}{\ell^2}\,sgn(z)^2
+\frac{8\alpha\kappa}{\ell}\delta(z)\bigg)
\partial_{\lambda}^2-\half\bigg(1-\frac{4\alpha\kappa}
{\ell^2}\,sgn(z)^2\bigg)\,\partial_z^2
+\frac{8\alpha\kappa}{\ell^2}sgn(z)\delta(z)\partial_z\nonumber\\ 
&~~~~~~+\frac{3}{2(\ell+|z|)}\bigg(1-\frac{4\alpha\kappa}
{\ell^2}\,sgn(z)^2\bigg)sgn(z)\partial_z\bigg]h_{\mu\nu}(x,z)=0\,.
\label{4dttgraviton}
\end{eqnarray}
With the GB term, $\ell$ admits two values
\footnote{$\ell_{\pm}^2=\frac{4\alpha'}{M_*^2}\bigg[1\pm
\sqrt{1+\frac{4\alpha'\Lambda}{3M_*^5}}\bigg]^{-1}=\ell^2$ and the physical 
case is explained by $\ell_-$ solution with $\Lambda<0$ and $\alpha'>0$. 
Here $[\Lambda]$ has mass dimension of $M^5$.}: 
$\ell_{\pm}$. 
Except the term with Dirac delta function, which vanishes for $|z|>0$ and 
$sgn(z)^2=1$, the equation~(\ref{4dttgraviton}) has a common factor 
$(1-4\alpha\kappa/{\ell^2})$. This indeed suggests that the microphysics 
on the brane with EGB terms will result in a renormalization of 
Einstein constant. This observation becomes clear when one evaluates 
the strength of gravity for a static source on the brane, or studies the 
behavior of graviton propagators.
     
To make a correspondence with the RS-type potential in $D=5$, we define the 
metric perturbation in canonically normalized form, i.e. 
$h_{\mu\nu}=e^{(D-2)\,A(z)/2}\tilde{h}_{\mu\nu}$, and look for the solutions 
of the form $\tilde{h}_{\mu\nu}(x,z)=\epsilon_{\mu\nu} e^{ip.x}\psi(z)$. Here 
$\epsilon_{\mu\nu}$ is the polarization tensor of the graviton wave 
function, $m^2=-p^2$ and $m(=\sqrt{- p.p})$ is the $5$-dimensional 
Kaluza-Klein (KK) mass of the gapless continuum modes. Then we obtain 
the following analog non-relativistic Schr\"odinger equation:
\begin{equation}
\bigg[-\partial_z^2+ V_{eff}(z)\bigg]\psi(z)= 
m^2\bigg[1+\frac{2\gamma\,\ell}{(1-\gamma)}\delta(z)\bigg]\psi(z)
\label{schrodinger}\,,
\end{equation}
where 
\begin{equation}
V_{eff}(z)=\frac{15}{4\big(|z|+\ell\big)^2}
-\frac{3}{\ell}\,\delta(z)\bigg[1-\frac{4\gamma\,\ell}{3(1-\gamma)}\,sgn(z)\,
\bigg(\partial_z+\frac{3}{2\ell}\, sgn(z)\bigg)\bigg]
\label{eff.poten}
\end{equation}
and $\gamma=4\alpha\kappa/{\ell^2}$. As is obvious, $\ell$ is the 
only dimensionful scale in the theory.
The attractive $\delta(z)$ term 
implies the existence of a localized zero-mode solution for the 
graviton wave function 
$h^{\mu\nu}_{(0)}=e^{2A(z)}\epsilon^{\mu\nu}e^{ip.x}$. The limit 
$\gamma\to 0$ corresponds to the RS potential. For 
$z\neq 0$ (or more precisely for $|z|>>\ell$), 
the eq.~(\ref{schrodinger}) reduces to 
\begin{equation}
\bigg[-\partial_z^2+ V(z)\bigg]\psi_m(z)= m^2\,\psi_m(z)
\label{eqn.tail}\,.
\end{equation}
The tail of the potential~(\ref{eff.poten}) is given by 
\begin{equation}
V(z)\sim \frac{\nu(\nu+1)}{z^2}=\frac{n\,(n+2)(n+4)}{4\,z_i^2}
\end{equation}
for a large $|z|$, and for the potential~(\ref{eff.poten}) we have 
$\nu=3/2$. Here $n$ counts the number of extra dimensions. This is 
the potential studied in two beautiful papers by Randall and 
Sundrum~\cite{RS} and Csaki, Erlich, Hollowood and Shirman~\cite{CEHS}. 
It is worth noting that for $|z|>>\ell$ the form of 
the potential~(\ref{eff.poten}) is the same as the RS potential. 
In fact, the eigenmodes (with $m^2\neq 0$) on the brane will be affected by 
GB curvature term, but gravity in the bulk will not be delocalized by 
higher-curvature terms in GB combination. The precise continuum mode for the 
potential~(\ref{eff.poten}) is given by a linear combination of the Bessel 
functions 
\begin{eqnarray}
\psi(z)\sim a\, (m(|z|+\ell))^{1/2}\, J_{2}\big(m(|z|+\ell)\big)
+b\,\big(m(|z|+\ell)\big)^{1/2}\,Y_{2}\big(m(|z|+\ell)\big)\,. 
\end{eqnarray}
The zero mode graviton 
wavefunction follows from $m\to 0$ limit of 
$m^2(|z|+ \ell)^{1/2} Y_2\big(m(|z|+ \ell)\big)$. When one substitutes the 
bound-state solution~\cite{KKL}     
\begin{equation}
\psi_0(z)\sim \ell\,\Big[a\,(|z|/\ell+1)^{-3/2}+ 
b\,(|z|/\ell+1)^{5/2}\Big]\label{eigenfn}
\end{equation}
into~(\ref{schrodinger}) and integrates in the neighborhood of $z=0$, 
admitting $sgn(z)^2=1$, one gets 
\begin{eqnarray}
3a (-1+2\gamma+1-2\gamma)
+5b(1-\gamma-2\gamma+\frac{3}{5}(1-3\gamma))=0\,,
\end{eqnarray}
which implies that $b=0$. Since the coefficient of the growing mode 
of the wave function vanishes, only massless graviton appears 
to be localized on the brane. This behavior is observed not because of the 
combination~(\ref{eigenfn}), rather due to the unique zero mode 
wavefunction. Indeed, for the case $m^2=0$ (massless graviton), the GB 
coupling does not change the boundary condition at all, independently 
whether $sgn(z)^2$ is zero or unity at the brane. For the normalizable 
zero-mode wave function $\psi_0(z)\sim \ell\, (|z|/\ell+1)^{-3/2}$, the 
boundary condition on the brane requires 
$(\partial_z+(3/2\ell)\,sgn(z))\psi(0)=0$. This implies that the product term 
$\delta(z)\,\big(\partial_z+(3/2\ell)\, sgn(z)\big)\, \psi_0(z)$ is 
precisely zero and hence $V_{eff}$ in~(\ref{eff.poten}) reduces to the RS 
potential. This suggests that one has to satisfy just the boundary condition 
implied by the $\delta$-function potential as in $\alpha=0$ case. However, 
for $m^2>0$ (small 
$m$ is relevant at long distances along the brane), the situation 
is significantly different. 
In the latter case, to satisfy the boundary condition implied by the 
$\delta$-function 
potential on the brane at $z=0$, one must choose the linear combination
\begin{equation}
\psi_m(z)\sim N_m \sqrt{|z|+\ell}\,\, \Big[Y_2\big(m(|z|+\ell)\big) 
+A\, J_2\big(m(|z|+\ell)\big)\Big]\,,
\end{equation}
where, 
\begin{eqnarray}
A=-\frac{Y_1(m\ell)+\chi\,m\ell\, 
Y_2 (m\ell)}{J_1\big(m\ell)+\chi\, m\ell\,
J_2 (m\ell)}\,,
\end{eqnarray}
where $\chi=\gamma/(1-\gamma)$. One has $A\equiv4/{(\pi \ell^2 m^2)}$ for 
$\gamma=0$. For a non-trivial 
$\gamma$ the graviton wavefunction corresponding to the continuum of KK 
modes is given by\footnote{Here $\psi_m(z)$ is to be normalized to unity 
(as plane waves) over a period at $|z|\to \infty$.} 
\begin{eqnarray}
\psi_m(0)&=& \sqrt{\frac{m\ell}{2}}\,
\frac{\big(J_1(m\ell)+\chi\,m\ell J_2(m\ell)\big) Y_2(m\ell)-
\big(Y_1(m\ell)+\chi\,m\ell\, Y_2(m\ell)\big) J_2(m\ell)}
{\sqrt{\big(J_1(m\ell)+\chi\,m\ell\, J_2(m\ell)\big)^2+
\big(Y_1(m\ell)+\chi\,m\ell\, Y_2(m\ell)\big)^2}}\nn\\
&\simeq& \sqrt{\frac{m\ell}{2}}\,\frac{1}{1+2\chi}\,,
\end{eqnarray}
where we have used $J_1(x) Y_2(x)-J_2(x) Y_1(x)=-2/(\pi x)$ and for 
small $m\ell (\equiv x)$,  
\begin{eqnarray} 
\sqrt{\big(J_1+\chi\, x\, J_2\big)^2+\big(Y_1+\chi\, x\, Y_2\big)^2}
&=&\frac{2(1+2\chi)}{\pi x}+\frac{x}{\pi}\,
\big[\log (x/2)+(\Gamma-1/2)-\chi\big]+\cdots\nn\\
&\approx &\frac{2(1+2\chi)}{\pi x}\,.
\end{eqnarray}
One can estimate the static gravitational potential at long 
distances along the brane due to the zero-mode contribution plus the 
gapless KK modes (the latter gives a leading order correction to 
potential). The 
four-dimensional gravitational potential due to a point source of mass 
$m_*$ localized on the brane can be estimated to be 
\begin{eqnarray}
U(|\bf r|)&\sim& - G_N\frac{m_*}{|\bf r|}
-(1-\gamma)^{-1}\,\frac{m_*}{M_*^3}\int_{m\neq 0}^{\infty}{}
dm\frac{e^{-m|\bf r|}}{|\bf r|} \psi_m(0)^2\nn\\ 
&\sim& -G_N\frac{m_*}{{\bf |r|}}
\bigg[1+\frac{1}{1+2\chi}\,\frac{\ell^2}{2 |\bf r|^2}+\cdots\bigg]
\label{pot.corr}\,,
\end{eqnarray}
where the factor $(1-\gamma)^{-1}$ arises from the linearized bulk 
equation, precisely, when one writes an expression valid on either side of 
the brane, and this contributes equally to zero-mode and KK mode, but one can 
still absorb such terms in $G_N$. 
Further $|\bf r|$ is the physical distance from the test particle along 
the brane. Here the modified Newton constant is defined by the relation 
\begin{equation}
(G_N)^{-1}\sim M_*^{3}\,\ell\,(1-\gamma) (1+2\chi)
\sim M_*^3\,\ell\,(1+\gamma)
\label{newtonconst}\,. 
\end{equation}
Eq.~(\ref{pot.corr}) is the result we verify by looking at the 
two-point Green function for one graviton exchange on the brane~\cite{CNW}. 
In the limit $\alpha\to 0$, one recovers the correction of 
the effective $4$-dimensional Newtonian potential due to RS and $M_{pl}^2=
M_*^3\ell$ in the standard way. The leading term due to the bound state is 
the usual Newtonian potential but with a modified Newton constant. As the 
$AdS$ curvature scale $\ell$ is expected in the order the Planck length (or 
fundamental string scale)\footnote{In fact, in the presence of 
Gauss-Bonnet term, if one fine tunes as 
$\Lambda \sim - 3 M_*^5/4\alpha'< 0 $, then one has 
$\ell\sim 2\sqrt{\alpha'} M_*^{-1}$, which for small 
$\alpha'$ can still be in the order of $M_{pl}^{-1}$.} and $r$ is the 
distance tested with gravity, for $r>>\ell$ the continuum KK modes 
with $m>>1/\ell$ are much suppressed, though there is an unsuppresed wave 
function in the region $r<<\ell$. Further, 
since $\gamma\equiv 4\alpha' M_*^{-2}\ell^{-2}$, 
for a large mass hierarchy between $M_*$ and $M_{pl}$, one needs 
$\ell M_*>1$. In the RS scenario, since $\ell M_*\sim 1$, $\gamma$ will be in 
the order of $\alpha'$ and small enough, but in the ADD approach the coupling 
$\gamma$ would be very small. In any case, to avoid anti-gravity 
effect, (i.e., for $G_N > 0$) one has to impose $1> \gamma > -1/2$.      

Obviously, in the higher dimensional theories $M_{pl}$ is determined by 
the extra space bulk curvature and the fundamental mass scale $M_*$. 
In a precise way, $M_{pl}^2\sim M_*^{2+n} \ell^n$, where $n$ counts the 
extra spaces. In general, in $D\geq 5$, small 
$\alpha'$ is consistent with the fact that 
the contribution of higher-curvature terms in low energy limit is almost 
negligible. Moreover, the ${\cal R}^n$ corrections and any 
non-linear correction to the linearized Einstein-Hilbert action are much 
suppressed along the brane, and this suppression redefines a 
normalized Newton constant. 

In $D=6$, the effective two dimensional analog Schr\"odinger equation can 
have the form\footnote{For the metric~(\ref{conmetric}), a general 
solution to warp factor $A(z_i)$ in $D=6$ takes a form 
$A(z_1, z_2)=\log\big(|z_1|/\ell_1+|z_2|/\ell_2+1\big)$, but, 
here we assume $\ell_1=\ell_2=\ell$ for simplicity.}
\begin{equation}
\bigg[-\partial_{z_i}^2+ V_{eff}(z_i)\bigg]\psi(z_i)= 
m^2\bigg[1+\frac{4 \gamma\,\ell}{(1-6\gamma)}\,\delta(z_i)\bigg]\psi(z_i)\,,
\label{schrodinger2}
\end{equation}
where 
\begin{eqnarray}
V_{eff}(z_i)&=&\frac{12}{\big(|z_i|+\ell\big)^2}
-\frac{4\,e^{-A}}{\ell}\delta(z_1)\bigg[1+\frac{\gamma\,\ell^2}{(1-6\gamma)}
\bigg\{e^{2A}\partial_{z_2}^2-\frac{e^{A}}{\ell}\Big(3 sgn(z_1)\,\partial_{z_1}
-sgn(z_2)\,\partial_{z_2}\Big)-\frac{6}{\ell^2}\bigg\}\bigg]\nn\\
&&~~~~~~~~~~~~~~~+\big( z_1\leftrightarrow z_2\big)
-2\,\delta(z_1)\,\delta(z_2)\,(1-6\gamma)^{-1}\,
\Big[40\gamma-\kappa\,V_1\Big]
\label{eff.poten2}
\end{eqnarray}
Here $z_i=z_1,\, z_2$. Evidently, the zero mode wavefunction is given by
\begin{equation}
\psi_0\sim \ell\,\big(|z_1|/\ell+|z_2|/\ell+1\big)^{-2}\,. 
\end{equation}
Then the eq.~(\ref{eff.poten2}), {\it via.} the terms in the brace 
$\{\,\}$, gives $-4\,\delta(z_2)$ 
plus a trivial contribution, and also implies the boundary conditions 
$\big(\partial_{z_i}+(2/\ell)\,sgn(z_i)\big)\,\psi(z_i=0)=0$. 
Similar term arises for $z_1\leftrightarrow z_2$, and these will be added to 
the last term. Then the resultant coefficient of $\delta(z_1) \delta(z_2)$ 
will give a fine tuning condition required between the GB coupling and the 
$3$-brane tension $(V_1)$, the latter guarantees a non-vanishing 
$3$-brane tension at the intersection of $4$-branes for a non-trivial 
$\alpha$, and in the expression above the fine tuning conditions for 
bulk cosmological constant $\Lambda$ and possible $4$-brane tensions 
$V_{z_i}$ do not show up in our linear treatment, which we have actually 
subtracted from $\delta G_{\mu\nu}$ and $\delta H_{\mu\nu}$ 
(see the Appendix).
\section{Discussion}
An extension of the RS singular $3$-brane model in $D\geq 5$ reveals some 
novel features of the brane world gravity in Einstein-Hilbert-Gauss-Bonnet 
theory. In the presence of GB interaction, we obtain a power law correction 
to Newtonian potential, the latter receives a dominant contribution 
from the KK mode for a small (and preferably positive) GB coupling $\alpha$. 
It appeared that the GB term does not affect the massless graviton mode and 
interestingly enough, such term in $D\geq 5$ gives more information about the 
boundary condition(s) on the brane(s). Possibly, in $D\geq 6$, the GB term 
can support the feasible singularities at 
the branes and allow a non-trivial $3$-brane tension. This could indeed be 
a meaningful realization of the RS model in $D>5$. It appears that a study of 
the full analysis of the gravitational fields in brane-localized EGB gravity 
in general $AdS_D$ backgrounds and exact non-linear analysis might be 
interesting to further explore the RS brane world scenario. Very recently, 
working along this line, a localized gravity on the intersection of two 
orthogonal non-solitonic and solitonic $4$-branes in $D=6$ has been realized 
in~\cite{JEK}, and a similar result was known in the case of the Einstein 
gravity~\cite{NSGN}. 
\section*{Addendum} 

In order to make the $\delta$-function regularization scheme explicit, 
one replaces in Eq.~(\ref{schrodinger}) $\gamma$ 
by $\frac{\gamma}{(D-4)}\,sgn(z)^2$, so for $D=5$ one replaces 
$\gamma$ by $\gamma\,sgn(z)^2$, and also uses 
$sgn(z)^2\delta(z)=\delta(z)/3$. Then Eq.~(\ref{schrodinger}) reads   
\begin{equation}
\bigg[- \partial_z^2+ V_{eff}(z)\bigg]\psi(z)= 
m^2\bigg[1-\gamma\,sgn(z)^2+2\gamma\,\ell\,e^{A(z)}\delta(z)\bigg]\psi(z)
\label{schrodinger1}\,,
\end{equation}
where 
\begin{equation} 
V_{eff}(z)=\frac{15}{4\big(|z|+\ell\big)^2}
-\frac{3}{\ell}\,\delta(z)\bigg[1-\frac{2\ell\gamma}{9}
\left(\partial_z\psi(0_+)+\frac{3}{2\ell}\,\psi(0_+)\right)\bigg]\,,
\label{eff.poten1}
\end{equation} 
where $\gamma=4\alpha'\, M_*^{-2}\,\ell^{-2}$. Similar analogy of the 
$\delta$-function regularization is applied to Eq.~(\ref{schrodinger2}), 
where one has to replace $\gamma$ by $\frac{\gamma}{2}\,
\big(sgn(z_1)^2+sgn(z_2)^2\big)$. 

\section*{Acknowledgements} 
The author would like to thank J. Erlich, A. Karch, Z. Kakushadze and 
Y. S. Myung for many fruitful correspondences and wish to thank S.-H. Moon 
and H. M. Lee for helpful discussion. The support 
from the BK-21 Initiative in Physics (SNU, School of Physics) for a research 
visit of author to the University of Waterloo is highly acknowledged.      
\renewcommand{\theequation}{A.\arabic{equation}}\setcounter{equation}{0}
\section*{Appendix A\,\, Metric expansion in $D=5$ with Gauss-Bonnet term}
The graviton equations derived by varying the action~(\ref{action}) w.r.to 
$g^{ab}$ take the form 
\begin{equation}
G_{ab}+ \kappa\,H_{ab}= \kappa\,T^{(0)}_{ab}+\frac{\kappa}{2}\,T^{(m)}_{ab}
\label{eqn.A1}
\end{equation} 
where $H_{ab}$, an analogue of the Einstein tensor stemmed from the 
GB term, is given by
\begin{eqnarray}
H_{ab}&=&-\frac{\alpha}{2}\,g_{ab}\, 
(R^2-4 R_{cd}R^{cd}+ R_{cdef}R^{cdef})\nn\\
&&+2\alpha \big[ R R_{ab}- 2 R_{acbd}R^{cd} +
 R_{acde}R_b\,^{cde}-2R_a\,^c R_{bc}\big]\,.
\end{eqnarray}
The linearized equations for~(\ref{eqn.A1}) read as 
$\delta\hat{G}_{ab}+\kappa \delta^{(1)}\hat{H}_{ab}
=\frac{\kappa}{2}\, T_{ab}^{(m)}$ 
where, 
$\delta \hat{G}_{ab}=\delta G_{ab}-\delta {T^{(0)}}_{ac}\, h_b\,^c$ and 
$\delta \hat{H}_{ab}= \delta H_{ab}- H_{ac}\, h_b\,^c$. 
The metric~(\ref{conmetric}) is more convenient to 
simplify the analysis of gravitational fluctuations in the 
background $g_{ab}={\bar g}_{ab}+{\bar h}_{ab}$. 
To make the analysis more general 
and easy to handle, we perform a conformal tranformation, 
$g_{ab}=e^{-2A(z)} {\tilde g}_{ab}$, and perturb the background metric as 
${\tilde g}_{ab}=\eta_{ab}+ h_{ab}$, so that the indices are raised or 
lowered with ${\tilde g}_{ab}$. 
For the metric fluctuations around the RS background in five dimensions 
$ds^2=e^{-2A(z)}\Big[\big(\eta_{\mu\nu}+h_{\mu\nu}\big) dx^\mu\,dx^\nu
+dz^2\Big]$, to the first order in fluctuations, we obtain 
\begin{eqnarray}
\delta^{(1)}\hat{G}_{\mu\nu}&=&-\frac{1}{2}
\partial_a\partial^a (h_{\mu\nu}-\eta_{\mu\nu}h)
+\frac{3}{2}\partial^zA\,\partial_z\big(h_{\mu\nu}
-\eta_{\mu\nu}h\big)\nn\\
& &+\half(2\partial_{(\mu}\partial^\lambda 
h_{\nu)\lambda}-\partial_\mu\partial_\nu h-\eta_{\mu\nu}
\partial_\lambda\partial_\rho h^{\lambda\rho})\label{gmunuhat},\\
\delta \hat{H}_{\mu\nu}&=&
2\alpha e^{2A}\bigg\{\big(-\partial_z\partial^zA\, \partial_\lambda^2
+\partial_z A\,\partial^z A\,\partial_z^2\big) 
(h_{\mu\nu}-\eta_{\mu\nu}h)\nn\\
&&+\big(2\partial_z\partial^zA-\partial_zA\,\partial^zA\big)
\partial^zA\, \partial_z h_{\mu\nu}
-\partial_z\partial^z A\,  \eta_{\mu\nu}\partial_\lambda\partial_\rho 
h^{\lambda\rho}\nn\\
& &+\big(\partial_z\partial^z A - 2 \partial_zA\, \partial^zA\big)
\big(2\partial_{(\mu}\partial^\lambda h_{\nu\lambda}-\partial_\mu
\partial_\nu h - \eta_{\mu\nu} \partial^zA\, \partial_z h\big)\bigg\}
\label{hmunu}\,,\\
\delta\hat{H}_{5\mu}&=&-4\alpha{A^\p}\,^2 e^{2A}\,\delta^{(1)}\hat{G}_{5\mu}= 
-2\alpha{A^\p}\,^2 e^{2A}\,\partial_z 
\big(\partial^\lambda h_{\mu\lambda}-\partial_\mu h\big)\,,\\
\delta \hat{H}_{55}&=&-4\alpha {A^\p}\,^2 e^{2A}
\delta^{(1)}\hat{G}_{55}= 2\alpha {A^\p}\,^2 e^{2A}\,
\big[\partial^\mu\partial^\nu (h_{\mu\nu}
-\eta_{\mu\nu}h)+3 A^\prime\partial_z h\big]\,,\\
\delta T_{ab}^{(0)}&=&-e^{-2A(z)}\Big[\Lambda h_{ab}+e^{A(z)}\,\delta_a^\mu
\,\delta_b^\nu\,\big(h_{\mu\nu}-\half\eta_{\mu\nu}\,h_i^i\big)
\delta(z)\,V(z)\Big]\,,\\
\delta T_{\mu\nu}&=&(-\Lambda - V \delta(z))
h_{\mu\nu}+T^{(m)}_{\mu\nu}\,,~~
\delta \hat{T}_{5\mu}=T^{(m)}_{5\mu},\,~~ 
\delta \hat{T}_{55}=T^{(m)}_{55}\label{t55}\,.
\end{eqnarray}

\end{document}